\documentclass[graphicx]{revtex4-1}

\usepackage{graphicx}
\draft 

\begin{document}


\title{Three-photon excitation of quantum dots with a telecom band ultrafast fiber laser} 



\author{M.J. Petrasiunas}
\email[]{m.petrasiunas@griffith.edu.au}
\author{J.B.O. Wood}
\author{D. Kielpinski}
\author{E.W. Streed}
\affiliation{Centre for Quantum Dynamics, Griffith University \\ 170 Kessels Road, Nathan, Queensland 4111, Australia}


\date{\today}

\begin{abstract}
We demonstrate three-photon excitation in quantum dots with a mode-locked fiber laser operating in the telecommunications band. We compare spectra and intensity dependence of fluorescence from one- and three-photon excitation of commercially available 640 nm quantum dots, using a 372 nm diode laser for one-photon excitation and 116 fs pulses from a mode-locked fiber laser with a center wavelength of 1575 nm for three-photon excitation.
\end{abstract}

\maketitle 

\section{Introduction}

Quantum dots are semiconductor nanocrystals, 1-20 nm in size \cite{Murray1993,Pietryga2004}, that possess a discrete energy level structure. The level structure is determined by the size, composition and shape of the quantum dot \cite{Empedocles1996,Norris1996}. Particular focus has been placed on II-VI semiconductors, such as CdSe \cite{Murray1993}. By selecting for size, the fluorescence emission can be tailored to the application, which is attractive for use in biological labels \cite{BruchezJr.1998,Alivisatos2004,Michalet2005,Resch-Genger2008}. In addition, the quantum dot surface can be chemically functionalized to bind to specific biomolecules. Compared to other fluorescent molecules, quantum dots possess high brightness, low bleaching rate and absorption over a wide wavelength range \cite{Michalet2005,Resch-Genger2008}. Similar to the fluorescence from other labels, quantum dot fluorescence can be driven by both single- and multiphoton processes \cite{Zipfel2003,Bentley2007,Lad2007,Chon2004,He2007}.

Fluorescent labels are conventionally excited by short wavelength light to produce fluorescence emissions. Ultraviolet (UV) light is often used with quantum dots due to their absorption that increases towards shorter wavelengths\cite{Resch-Genger2008}. However UV is not ideal for use in living cells, due to strong absorption and scattering of UV light by surrounding biological tissue. The use of shorter wavelengths leads to a higher potential for photochemical reactions and an increased possibility for the decomposition and subsequent release of component heavy metals, such as cadmium (Cd) \cite{Derfus2004,Hardman2006}. Through excitation by multiphoton processes, it is possible to avoid the unwanted effects from exposure to shorter wavelengths in biological applications. Two-photon excitation of quantum dots is now routinely performed, often using Ti:sapphire lasers operating around 700-1000 nm \cite{He2007,Dahan2001}.

For imaging applications, it is often important for the excitation light to penetrate far into the sample, but the penetration depth is limited by Rayleigh scattering due to the effect on beam quality. Since Rayleigh scattering scales as $\lambda^{-4}$, long-wavelength multiphoton excitation seems promising for bioimaging applications. In a recent study, Horton et al. \cite{Horton2013} demonstrated in-vivo three-photon microscopy with an Er-doped fiber system and red fluorescent protein labels. They found that an optimum wavelength window due to scattering and absorption in tissue lies close to the telecom band. Quantum dots are attractive for use in such applications due to their large three-photon absorption cross-section \cite{Bentley2007,Lad2007}.

We demonstrate three-photon excitation and fluorescence in commercially available CdSe quantum dots using a custom built mode-locked Er-doped fiber laser (EDFL). Such lasers, and their associated optical components, are central to telecommunications and have therefore been the subject of extensive industrial research. These developments mean that they provide a stable, versatile and comparatively inexpensive method to generate ultrashort pulses for three-photon spectroscopy of quantum dots. We also excited single-photon fluorescence using a 372 nm diode laser to compare one- and three-photon fluorescence emission spectra, excitation intensity dependence and bleaching rate. Previous investigations have also demonstrated two- and three-photon excitation in quantum dots \cite{Chon2004,He2007}. However, to our knowledge, multiphoton excitation and absorption have not yet been investigated in quantum dots for excitation wavelengths in the telecom band.

The wavelength of expected fluorescence emissions from the sample is determined by the quantum dot energy level structure, which can be simplified to a ground state, fluorescent excited state and a continuum of absorption states \cite{Alivisatos1996,Reimann2002}. Upon excitation to this continuum of states, a quantum dot undergoes a non-radiative decay to the excited state, followed by fluorescent decay back to the ground state\cite{Alivisatos1996}. In addition, intense excitation pulses will interact nonlinearly with the sample to generate harmonics of the excitation wavelength. In quantum dots, the third-harmonic generation (3HG) process is expected to be the most significant of these \cite{Dharmadhikari1999}. The process of second harmonic generation (SHG) is negligible in comparison to 3HG due to the centrosymmetric crystal structure of CdSe \cite{Banyai1993,Feng2006}. As the product of a coherent process, the harmonics are emitted in line with the excitation beam.

\section{Method}

\begin{figure}
\includegraphics[width=7cm]{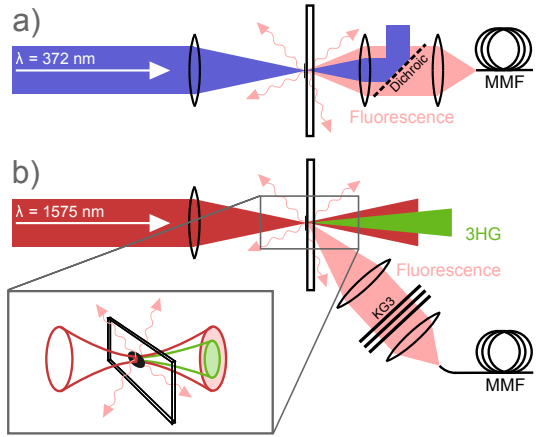}%
\caption{\label{fig1-setup}
Experimental Setup. Schematic for spectroscopy of quantum dots. (a) 372 nm light is focused onto the quantum dot sample and fluorescence is collected and coupled into a multimode fiber. Fluorescence is separated from residual excitation light using a Semrock FF511-Di01 dichroic mirror. (b) 1575 nm pulses are focused onto the quantum dot sample and fluorescence is collected off-axis to avoid excessive coupling of third harmonic (3HG) light. The majority of residual excitation and 3HG light continues to propagate on-axis past the slide and is not coupled into the multimode fiber (MMF). As some scattered light is still collected, KG3 coloured glass filters are used to block any remaining IR.}%
\end{figure}

Figure \ref{fig1-setup} shows the experimental setup used to probe the quantum dots. Colloidal Invitrogen\texttrademark{} Lumidot\texttrademark{} CdSe/ZnS 640 core-shell quantum dots, with a nominal fluorescence wavelength of 640 nm, were deposited on a sample slide. The toluene solvent was evaporated off, leaving a static quantum dot sample which was placed at the focus of the excitation beam. Collected fluorescence was coupled into a multimode fiber, where the spectrum was measured by a fiber-coupled spectrometer.

For the three-photon spectroscopy of the quantum dot sample, a mode-locked EDFL was used to produce 1564 nm pulses at a repetition rate of 300 MHz. 116 fs pulses with a maximum average power of 186 mW were produced through Er-doped fiber amplification and soliton compression. The laser was focused to a spot with a $14.1\pm0.2\;\textrm{\ensuremath{\mu}m}$ $1/e^2$ radius, yielding an average intensity up to $48.8\pm0.5\;\textrm{kW/cm}^{2}$. The soliton self-frequency shift (SSFS) \cite{Mitschke1986,Gordon1986} of the amplified pulses shifted the center wavelength to 1575 nm, increasing the spectral bandwidth to 85 nm. The pulse durations were measured by frequency resolved optical gating (FROG) using a MesaPhotonics FROGScan with an Ocean Optics HR2000+ spectrometer, and a wavelength range of 700-881 nm. The spectrum was measured using a HP 70950A optical spectrum analyser.

The resulting fluorescence emitted from the quantum dots was collected off-axis with a 35 mm focal length lens to spatially separate it from the large amount of light produced by 3HG. However, some of the 3HG and excitation light was scattered by the sample and collected. The fiber-coupled fluorescence was detected by an Ocean Optics USB650 spectrometer. Three 2 mm thick KG3 coloured-glass filters were used to block any remaining IR, with a calculated attenuation of $-80$ dB, in order to prevent damage to the detector.

The fluorescence detection spectrometer was used to capture all data for the fluorescence and 3HG spectra, and its dependence on the intensity of the excitation beam and exposure time.  Measurements of the spectra were taken for different input intensities using a variable ND filter, and integration times from 5-30 s were used. The relative spectrum amplitudes were inferred from Gaussian fit profiles of the spectra.

To check that the quantum dots behaved as expected under UV excitation, we applied up to 3.5 mW of 372 nm light from a fiber-coupled Nichia diode laser to the quantum dot sample. Basic optical isolation was employed, using a quarter waveplate and linear polariser, to ensure that the operation of the diode laser was not compromised by optical feedback from the single mode fiber. The UV light was focused to a spot with a $5.6\;\textrm{\ensuremath{\mu}m}$ $1/e^2$ radius at the quantum dot sample, with a maximum intensity of $3.54\;\textrm{kW/cm}^{2}$.

The fluorescence excited by the UV laser was collected by an $\textrm{NA}=0.60$ aspheric lens with a 4.02 mm focal length. This was separated from the residual excitation light by a Semrock FF511-Di01 dichroic mirror, with an edge wavelength of 511 nm, and coupled into MMF. An Ocean Optics QE65000 spectrometer, with a 316-1021 nm wavelength range and 0.22 nm/pixel resolution, was used to measure the spectrum of the UV-induced fluorescence. Measurements were taken with varying input intensities and an integration time of 10 ms.

\section{Spectra}

\begin{figure}
\includegraphics[width=7cm]{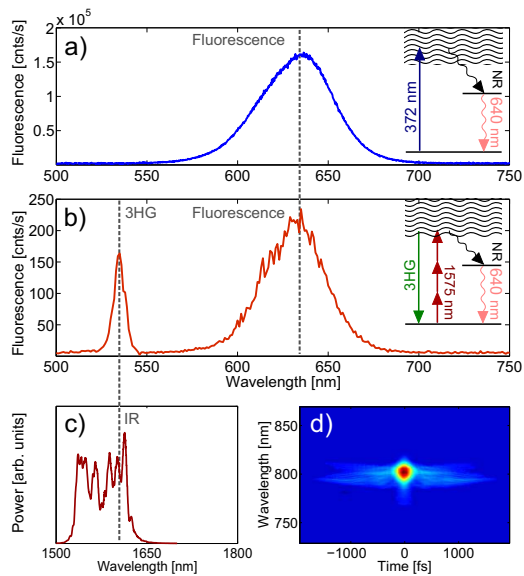}%
\caption{\label{fig2-spectra}
Comparison of spectra from (a) $\lambda$=372 nm excitation and (b) $\lambda$=1575 nm excitation, shown with the energy level structures and excitation pathways. In (b), both the 640 nm fluorescence peak and a 540 nm 3HG peak are observed. (c) Spectrum of IR excitation beam on a linear scale, where the wavelength axis aligns at a ratio of 3:1 to that of (b), showing the correlation of the 3HG peak with the fundamental. (d) FROG trace of IR pulse, where the wavelength axis refers to the second harmonic of the excitation wavelength.}%
\end{figure}

Figure \ref{fig2-spectra} shows the spectra from UV and IR excitation. From a Gaussian fit of the fluorescence spectra, the center wavelength from linear excitation is $632.6\pm0.1\;\textrm{nm}$, where it is $631.4\pm0.2\;\textrm{nm}$ in the spectrum produced by three-photon excitation. The FWHM linewidths are determined to be $47.8\pm0.1\;\textrm{nm}$ and $40.7\pm0.4\;\textrm{nm}$, respectively. The near-identical characteristics of the fluorescence spectra under UV and IR excitation indicate that the same excited energy band is reached in both cases.

In the three-photon excitation spectrum, the secondary peak at $534.8\pm0.1\;\textrm{nm}$ corresponds with the third harmonic of the IR excitation pulse. The peak is visible in the off-axis configuration due to scattering from the quantum dot sample. As shown in Figure \ref{fig2-spectra}, the third harmonic is generated predominantly from the red-shifted soliton at the long-wavelength end of the excitation pulse spectrum. This part of the excitation spectrum, according to FROG trace measurements, corresponds to the most intense part of the excitation pulse.

As a static sample of quantum dots were used for this experiment, the time dependent spectra were also observed in each case, using an integration time of 500 ms over 10 minutes, to compare bleaching rates of the linear and three-photon excitation processes. We observed that although a higher excitation intensity was used for the three-photon process, the emission signal of the quantum dots was bleached to 50\% of the starting intensity after 566 s, compared to 103 s under linear excitation. However the bleaching of the sample under IR excitation was not fit well by the expected double-exponential profile, suggesting that it may not follow a conventional bleaching mechanism. Further study is needed to properly assess the meaning of these preliminary observations.

\section{Intensity dependence}

\begin{figure}
\includegraphics[width=7.5cm]{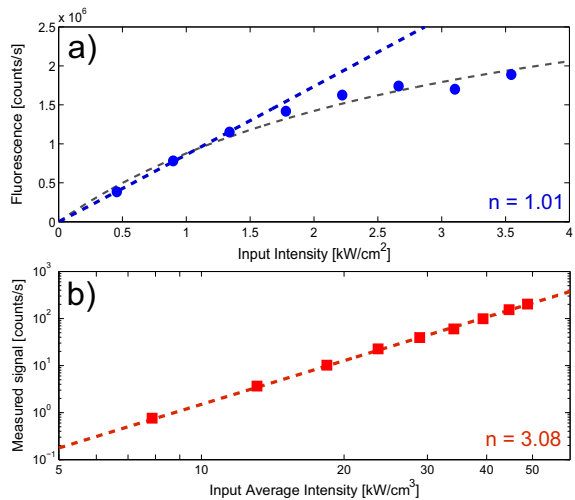}%
\caption{\label{fig3-power}
Fluorescence rate dependence of one- and three-photon excitation. (a) Intensity dependent fluorescence measured under UV excitation shows a saturation intensity of $3.3\pm1.1\;\textrm{kW/cm}^{2}$ from the saturation fit, given by the dashed gray line. Excluding points above 50\% of the saturation intensity, and fitting a power law $I^{n}$, gives $n=1.01\pm0.15$ (dashed blue line). (b) A power law fit of the intensity dependent fluorescence measured under three-photon excitation is shown, with an index $n=3.08\pm0.06$ (dashed red line).}%
\end{figure}

To verify the photon orders of the two excitation processes and search for saturation of the excitation, we measured the dependence of the fluorescence on excitation intensity, as shown in Figure \ref{fig3-power}. The fluorescence spectra for both the one- and three-photon measurements, taken at varying excitation intensities, were fit with Gaussian profiles. The amplitudes of the fit profiles were compared using power law fits, as shown in Figure \ref{fig3-power}, in order to determine a $F \propto I^{n}$ power law dependence between excitation intensity $I$ and fluorescence counts $F$, for each of the processes. First- and third-order dependencies are expected for the one- and three-photon processes, respectively. Saturation of the quantum dots is observed under UV excitation, with saturation intensity $I_{sat}=3.3\pm1.1\;\textrm{kW/cm}^{2}$, found using a fit of the form:
\begin{equation}
F\propto\frac{s}{1+s} \qquad s=I/I_{sat}
\end{equation}
 By excluding data points greater than 50\% of the saturation intensity found for the UV excitation data, the index of the UV intensity dependence was found to be $1.01\pm0.15$. The index for the IR induced fluorescence was $3.08\pm0.06$, confirming three-photon behaviour. Third harmonic intensity was also found to exhibit cubic behaviour. As such, the fluorescence under IR excitation could either occur via a direct three-photon mechanism or via absorption of a third harmonic photon. This experiment does not distinguish between the two mechanisms.

\section{Discussion}

Our results show three-photon excitation and fluorescence in quantum dots, with the aid of ultrafast fiber lasers. With this comes the potential to improve the effectiveness of quantum dot based biological label schemes, using low-cost hardware. Telecom-band fiber lasers provide an economical source of laser hardware that has the potential to supplant more conventional laser hardware such as Ti:Sapphire in many applications - of which biomicroscopy applications are no exception.

The experimental data for three-photon excitation shows that there are no significant changes or shifts in the fluorescence spectrum compared to the linear excitation case, and the third order dependence on excitation power shows that the quantum dot sample is indeed excited through a three-photon process. The process in the study by Horton et al. \cite{Horton2013} could thus be replicated using quantum dots as a more favourable label agent \cite{Michalet2005,Resch-Genger2008,Bentley2007,Lad2007}.

In future work, we aim to address the implementation of quantum dots in biological imaging within tissue, using multiphoton processes to maximize viewing depth. Further work could improve the design of the fiber laser system used for this process and use additional types of quantum dots - including more biocompatible materials such as InP and different fluorescence wavelengths. Optimising quantum dots for longer wavelength fluorescence will significantly reduce scattering of the fluorescence signal, once again increasing viewing depth. By opting for a lower repetition rate, thus increasing the pulse energy of the fiber laser, it will also be possible to increase the multiphoton fluorescence signal of the labels while reducing chances of cell damage due to high average power levels.

\section{Conclusion}

We provide the first demonstration of three-photon excitation in quantum dots using a fiber laser operating in the telecommunications band. The experiment was conducted using a mode locked Er-doped fiber laser producing 116 fs pulses with a center wavelength of 1575 nm. We compare results from one- and three-photon excitation experiments to confirm that the three-photon spectrum is similar, and to demonstrate the third order intensity dependence of the process as an additional means of verification. We found that the fluorescence intensity scaled with input excitation power as expected for a three-photon process. Three-photon excitation in quantum dots using ultrafast fiber lasers provides an important step toward more effective biological imaging by minimizing scattering of the excitation light and increasing the possible penetration depth.

\begin{acknowledgments}

D.K. was supported by an Australian Research Council Future Fellowship (FT110100513). We would like to acknowledge Prof. Jay L. Nadeau of McGill University for sparking our interest in the area.

\end{acknowledgments}

\end{document}